\documentstyle[preprint,aps]{revtex}
\tightenlines
\newcommand{\Fsc}{{\cal F}}
\begin{document}
\draft
\title{Spontaneous DC Current Generation in a Resistively Shunted
Semiconductor Superlattice Driven by a TeraHertz Field}
\author{Kirill N. Alekseev$^{1,2,3}$,
Ethan H. Cannon$^1$, Jonathan C. McKinney$^1$,
Feodor V. Kusmartsev$^{2,4}$\cite{fedya},
and David K. Campbell$^1$ }
\address{$^1$Department of Physics,
University of Illinois at Urbana-Champaign,
1110 West Green St.,
Urbana, IL 61801\\
$^2$NORDITA, Bledgamsvej 17, DK-2100, Copenhagen 0,
Denmark\\
$^3$Theory of Nonlinear Processes Laboratory,
Kirensky Institute of Physics,
Krasnoyarsk 660036, Russia\\
$^4$Department of Physics,
      Loughborough University,
      Loughborough
      LE11 3TU, UK}
\maketitle
\begin{abstract}
We study a resistively shunted semiconductor superlattice subject
to a high-frequency electric field. Using a balance equation approach
that incorporates the influence of the electric circuit,
we determine numerically a range of amplitude and frequency of the ac
field for which a dc bias and current are generated {\it
spontaneously} and show that this region is likely accessible to current
experiments. Our simulations reveal that the Bloch frequency 
corresponding to the
spontaneous dc bias is approximately an integer multiple of the 
ac field frequency.
\end{abstract}
In 1928 Bloch demonstrated \cite{bloch} that electrons in a
periodic lattice potential with period $a$ subject to a dc electric
field $E_{dc}$ undergo oscillations with characteristic frequency
$\omega_B=e a E_{dc}/\hbar$. In naturally occurring solids,
observations of Bloch oscillations are precluded by the
requirements of the extremely high applied electric fields
needed to reduce the Bloch oscillation period below the
dephasing times arising from the ever-present scattering due to impurities and
phonons. In 1970, however, Esaki and Tsu \cite{esaki}
recognized that the inherently longer spatial periods available
in semiconductor superlattices (SSLs) allow
Bloch oscillations in these artificial structures
to lie the the terahertz (THz) domain---comparable to or higher
than the corresponding scattering frequencies---even for modest
field strengths ($\sim$1 kV/cm). Thus, not
only should the Bloch oscillations be observable in SSLs,
but SSLs could also serve as devices to produce THz
radiation \cite{esaki}.

Since 1970, an enormous literature has evolved describing novel
physical phenomena in SSLs and their potential applications for
devices \cite{reviews}.
Among the most recent developments, which have relied
on advances in both electromagnetic radiation sources and
coupling techniques, are detailed observations of
the influence of a THz field on the dc conductivity of an SSL
\cite{guimaraes,ignatov1,keay,unterrainer}, including (1)
THz multiphoton-assisted tunneling
\cite{guimaraes}; (2)
ac-field-induced reduction of the dc current
\cite{ignatov1}; (3)
{\it absolute} negative conductance \cite{keay}; and (4)
resonant changes in
conductivity \cite{unterrainer}.

In the present article, we examine a question that is
roughly
the converse of the Bloch oscillations in an SSL:
namely, can a purely ac external field applied to an
appropriately
configured (but unbiased) SSL create a dc bias and
corresponding dc current?
In technical terms, this corresponds to a spontaneous
breaking of
spatial reflection symmetry.  We show that the answer to this 
question is affirmative.  Using the methods of nonlinear dynamics,
we find the condition for spontaneous dc current generation
in the circuit
consisting of the SSL shunted by an external resistance.
We observe that the Bloch frequency corresponding to the
spontaneously generated bias across the SSL is
approximately an integer multiple of the frequency of the external ac field.
Finally, we use the time-periodicity of the applied field to
understand the origin of this ``phase-locking''.

We consider an SSL of spatial period $a$, length $l$ and
area $S$ that is
homogeneously doped with $N$ carriers per unit volume. The
SSL is shunted by an external
measuring device of resistance $R$ and is exposed to an
external ac electric
field $E_{ext}(t)=E_0\cos\Omega t$.
We use the standard tight-binding dispersion relation for
electrons belonging
to a single miniband, so that
$\varepsilon(p)=\Delta/2 [1-\cos(pa/\hbar)]$,
with $\varepsilon$ being the electron energy, $p$ the quasimomentum
along the SSL's axis, and $\Delta$ the miniband width. We
describe the dynamics of electrons
by the balance equations  \cite{ignatov3,alekseev}
\begin{equation}
\dot{V}=-e E_{tot}(t)/m({\cal E}) -\gamma_v V,\quad
\dot{\cal E}=-e E_{tot}(t) V
-\gamma_\varepsilon ({\cal E}-{\cal E}^{(0)}),\quad
\dot{E_{sc}}=4\pi eNV/\epsilon_0-\alpha E_{sc}
\label{1}
\end{equation}
where $V(t)$ and ${\cal E}(t)$ are the electron velocity
and energy averaged
over the time-dependent distribution function satisfying 
the Boltzmann equation,
and $\gamma_v$ and $\gamma_{\varepsilon}$ are
phenomenological relaxation constants
for the average velocity and energy, respectively.
Elastic scattering by impurities, interface roughness and
structural disorder contributes to $\gamma_v$; inelastic phonon
scattering comprises the main channel of energy dissipation.
The parameter ${\cal E}^{(0)}$ describes the electron
equilibrium energy \cite{ignatov3}, which is a function of
the lattice temperature.
Also appearing in (\ref{1}) is the energy-dependent
effective mass $m({\cal E})=m_0 (1-2{\cal E}/\Delta)^{-1}$, where
$m_0\equiv\frac{2\hbar^2}{\Delta a^2}$.

The electric field acting on the electrons,
$E_{tot}(t)=E_{sc}(t)+E_{ext}(t)$,
is the sum of the external ac field $E_{ext}(t)$
and the self-consistent field $E_{sc}(t)$, which incorporates the
influence of the circuit and of the other electrons on a
single electron's dynamics. The phenomenological constant $\alpha$ describes
the relaxation of the self-consistent field \cite{alekseev}.
The self-consistent field is related to the voltage $U$
across the SSL by $E_{sc}=U/l$.
The current density through the SSL consists of two parts:
the displacement current
$j_{disp}=\frac{\epsilon_0}{4\pi}{\dot{E}}_{sc}$
($\epsilon_0$ is the average dielectric constant for the SSL)
and the current of ballistic electrons $j=-e N V$, where $V$ is the
solution to Eqs (\ref{1}).
Kirchoff's equation for the resistively shunted SSL,
$((\epsilon_0/4\pi)\dot{E}_{sc} - e N V) S +(E_{sc} l)/R=0$,
provides the circuit's contribution to the damping of the
self-consistent field,
$\alpha_{cir}=(RC)^{-1}$, $C=\frac{\epsilon_0 S}{4\pi l}$;
the total damping rate is the sum of
the rates from the external circuit and internal mechanisms.
Introducing the scaled variables $\Fsc(t)=e a E_{sc}(t)/\hbar$,
$v=m_0 V a/\hbar$, and $w=({\cal E}-\Delta/2)(\Delta/2)^{-1}$,
for which the lower (upper) edge
of the miniband corresponds to $w=-1$ ($w=+1$),
we obtain the self-consistent set of equations
\begin{equation}
\dot{v}=(\Fsc(t)+\omega_s\cos\Omega t) w -\gamma_v v,\quad
\dot{w}=-(\Fsc(t)+\omega_s\cos\Omega t) v -
\gamma_\varepsilon (w-w^{(0)}), \quad 
\dot{\Fsc}=\omega_c^2 v -\alpha \Fsc.
\label{3}
\end{equation}
%
Here $\omega_s=(e a E_0)/\hbar$ and
$\omega_c=[(2\pi e^2 N a^2\Delta)/(\hbar^2\epsilon_0)]^{1/2}$,
where $\omega_c$ is a characteristic frequency that can
be interpreted as a miniband plasma frequency \cite{alekseev}.
Throughout this paper
we assume that initially the electrons are at the bottom of
miniband, $w(0)=-1$,
with both current and voltage absent, $v(0)=\Fsc(0)=0$,
and that the electrons relax to the bottom of the miniband,
{\it i. e.} $w^{(0)}=-1$, where
$w^{(0)}=({\cal E}^{(0)}-\Delta/2)(\Delta/2)^{-1}$.

Since we are here interested primarily in stationary
solutions of (\ref{3}),
we do not consider transient effects. The appearance of a
dc current and
a dc bias indicates the presence of the zero-$^{th}$
Fourier harmonic in the
stationary solutions for $v(t)$ and $\Fsc(t)$.
Hence, we first establish how the existence of these 
zero-$^{th}$ harmonics
relates to the symmetry properties of (\ref{3}).  Note that the
appearance of these zero-$^{th}$ harmonics is accompanied
by other (normally forbidden) even harmonics, {\it e.g.} the second
harmonic.

The following transformation leaves the system (\ref{3})
invariant:
\begin{equation}
v\rightarrow -v,\quad \Fsc\rightarrow -\Fsc,\quad
t\rightarrow t+T_{sym}/2,\quad T_{sym}\equiv (2n+1)T,
\label{4}
\end{equation}
where $n$ is an integer and $T=2\pi/\Omega$ is the period
of the external ac field.
In other words, for any solution of system (\ref{3})
$\{v(t), w(t), \Fsc(t)\}$,
there exists a symmetry-related ``conjugate'' solution
$\{-v(t+T_{sym}/2), w(t+T_{sym}/2),-\Fsc(t+T_{sym}/2)\}$.
Physically, transformation (\ref{4}) demonstrates that both
signs of electron velocity $v$ and self-consistent field (voltage)
$\Fsc$ are equivalent in an SSL driven by an ac field without bias;
in contrast, the lower and upper
edges of the miniband are not equivalent, thus
$w\not\rightarrow -w$.

Importantly, the system (\ref{3}) possesses two
well-studied limiting cases \cite{alekseev}.
First, when $\gamma_v=\gamma_\varepsilon=0$, it can be
transformed to the
resistively shunted junction (RSJ) model familiar from the theory of
Josephson junctions \cite{d'humieres}.
Second, in the absence of an applied field
($\omega_s=0$), but for arbitrary damping,
it can be shown \cite{alekseev} to be equivalent to the
familiar Lorenz model \cite{aizawa}.
In both these limiting cases, symmetry properties have been
applied effectively to characterize and interpret solutions, and we
therefore expect this to prove true in the present, more general case.

The form of the symmetry transformation suggests the use of
a stroboscopic map of period $T/2$ to characterize the solutions of
system (\ref{3}):  This amounts to considering $v$, $w$,
$\Fsc$ at discrete times $t_n=n T/2$, where n is an integer.
We are especially interested in the projection of this
stroboscopic phase portrait onto the $v-\Fsc$ plane.
Let $\phi(t)$ be one of the variables $v(t)$ or $\Fsc(t)$
and denote by
$\langle\phi\rangle$ the value of $\phi$ averaged over a
large time interval.
By analogy to the RSJ model \cite{d'humieres,mcdonald}
and the Lorenz model with periodic external driving \cite{aizawa},
we have the following natural classification for attractors of
(\ref{3}) based on their symmetry properties:\\
(i) A {\bf symmetric limit cycle} is a solution that obeys the
equality $-\phi(t+T_{sym}/2)=\phi(t)$, which implies that it
is periodic with period $T_{sym}=(2n+1)T$;
as this period is $(2n+1)$ times the external driving period,
this limit cycle is also known as a period $(2n+1)$-solution.
The $T/2$-stroboscopic plot consists
of $2(2n+1)$ points, and its projection on the $v-\Fsc$
plane has inversion symmetry about the origin.
More importantly, $\langle\phi\rangle=0$, and therefore
{\it dc current and bias are absent}.
The simplest example of a symmetric limit cycle is a
period-1 solution of the form
$\phi(t)=A_{\phi}\cos\Omega t +B_{\phi}\sin \Omega t$,
where $A_{\phi}$ and $B_{\phi}$ are constants.\\
(ii) For a {\bf symmetry-broken limit cycle}, the $T/2$-stroboscopic
phase portrait in the $v-\Fsc$ plane lacks inversion symmetry
about the origin.
The simplest example is a period-1 solution of the form
$\phi(t)=A_{\phi}\cos\Omega t +B_{\phi}\sin \Omega t + C_{\phi}/2$
with zero harmonic equal to $C_{\phi}$ ($A_{\phi}$,
$B_{\phi}$ and
$C_{\phi}$ are constants). This solution satisfies the equality
$-\phi(t+T_{sym}/2)=\phi(t) - C_{\phi}$.
Actually, there are two symmetry-related conjugate limit cycles
differing from each other only by the sign of the constant
$C_{\phi}$, and
for fixed parameters, a trajectory selects one limit
cycle depending on its initial conditions.
For our case with fixed initial conditions,
a trajectory can switch to a limit cycle with different
symmetry only when the parameters are varied.
This means that
{\it for symmetry-broken limit cycles dc current and bias are nonzero}
and their signs are fixed for fixed values of the system
parameters.\\
(iii) For {\bf symmetric chaos}, the solutions are aperiodic, and
the $T/2$-stroboscopic phase portrait
has a fractal structure with an infinite number of points
but possesses inversion symmetry about the origin of
coordinates in the $v-\Fsc$ plane.
Such solutions correspond to strange attractors with
$\langle\phi\rangle=0$, and {\it no dc current}.\\
(iv) For {\bf asymmetric chaos}, the fractal pattern in the $T/2$
stroboscopic representation {\it lacks} this $v-\Fsc$ inversion
symmetry, and one has a symmetry-related
conjugate pair of strange attractors.
For asymmetric chaos,
$\langle\phi\rangle\not=0$,
and {\it dc current is spontaneously generated}.

We observe all four types of attractors in our model.
The most unusual type of behavior -- asymmetric chaos -- is
illustrated in Fig.~\ref{fig1}, which shows a stroboscopic phase
portrait of an attractor lacking inversion symmetry about the origin.
The symmetry-related conjugate attractor exists but is {\it
not} depicted in this Figure.

The above symmetry-based classification of attractors
demonstrates that stationary solutions both with and
without dc current can be realized in our model.
To determine where in parameter space the different types
of attractors exist, we have undertaken numerical simulation of system 
(\ref{3}) holding the dissipation rates fixed but varying the
external field strength and frequency.
For modern SSLs, the velocity relaxation times range from
tens of femtoseconds up to picoseconds.
If we take $\gamma_v^{-1}=0.35$ ps
\cite{unterrainer}, then at $\Delta=22$ meV, $a=42$ nm,
$N=1.2\times 10^{16}$ cm$^{-3}$ and $\epsilon_0\approx 13$ ($GaAs$),
we have $\gamma_v/\omega_c=0.1$. As a rule, the relaxation
time for energy is
an order of magnitude longer than for velocity
\cite{minot}, therefore we take
$\gamma_\varepsilon/\omega_c=0.01$. We choose
$\alpha/\omega_c=0.01$, which corresponds
to a resistance of $R=5.4$ kOhm and an SSL self-capacitance of
$C=0.64$ fF
($S=10^{-7}$ cm$^2$, $l=40\times a=1.68\times 10^{-4}$ cm).
For these parameters, the locations of
broken-symmetry attractors
are shown in the $\Omega$--$\omega_s$ plane in
Fig.~\ref{fig2}.
To distinguish genuine symmetry-breaking from long
transients and rounding errors, we take advantage of the
observed ``phase-locking'' and mark as symmetry-broken those
solutions having $\mid\langle\Fsc\rangle\mid>\Omega/4$.
There exists a small region of asymmetric chaos with
$\langle v\rangle$ and $\langle\Fsc\rangle\not=0$.
However, the majority of the symmetry-broken attractors are
symmetry-broken
limit cycles with higher even and odd harmonics as well as
the zero-$^{th}$ harmonic
\cite{remark}.
We also observe some solutions with subharmonics, 
corresponding to period doubling.

The dependence of the spontaneously
generated bias and dc current on the frequency of the external
field is indicated in Fig.~\ref{fig3}.
Recall that $\langle\Fsc\rangle$ is the {\it long-time average}
of the {\it scaled} self-consistent field, {\it i.e.} it reflects
the dc component of the spontaneously generated electric field
in the SSL and is in fact equal to the ``induced'' Bloch frequency
($\omega_B=e a E_{dc}/\hbar=\langle\Fsc\rangle$).
We observe from Fig.~\ref{fig3} that $\langle\Fsc\rangle$
is approximately equal to an integer multiple of $\Omega$.
This ``mode-locking'' is the analog of the well-known
zero-bias ac Josephson steps
in a Josephson junction \cite{d'humieres,note}.
But in contrast to the RSJ model, the ``mode-locking'' in
our case is only approximate -- the exact value of $\langle\Fsc\rangle$
depends weakly on the field strength $\omega_s$.
We observe ``mode-locking'' for periodic limit cycles as well as for
asymmetric chaos. The efficiency of ac to dc field conversion, {\it
i.e.} the ratio $\omega_B/\omega_s=\mid\! E_{dc}/E_0\!\mid$, is often
in the range $0.5$-$0.67$,
and the maximal efficiency we observed is $\approx 0.9$.
Therefore, a resistively shunted semiconductor superlattice
could be an effective THz ac field to dc current converter.

From (\ref{3}) follows that for {\it periodic} solutions
$\langle v\rangle=\alpha\omega_c^{-2}\langle\Fsc\rangle$.
Therefore, {\it if the bias is ``phase-locked''}
$\langle\Fsc\rangle\approx n \Omega$ ($n$ is an integer),
{\it then the dc current is also proportional to} $n\Omega$.
Although this formula does not apply to aperiodic
motion,  we observed the same dependence for asymmetric chaos,
at least during our simulation time of
$17\times 10^3$ $\omega_c^{-1}$.
To understand the origin of this ``phase-locking'', we
recall that the wave functions of a time-periodic Hamiltonian with
period $T$ can be written as Floquet functions, where
$\psi(x,t)=exp(-\imath \epsilon t/\hbar)u_\psi(x,t)$,
$\epsilon$ is the quasienergy,
and $u_\psi$ is periodic in time with period $T$ \cite{zak}.
Further, as shown by Krieger and Iafrate \cite{krieger},
when an electric field is applied to a perfect lattice,
the Bloch wave functions evolve as Houston functions
until tunneling to higher bands occurs.  Therefore, in
our single-band model {\it in the absence of scattering}
the electrons would evolve as Houston functions indefinitely.
But Zak has demonstrated \cite{zak} that for a Houston
function to be a Floquet
state, the dc component of the electric field must satisfy the
relation $eE_{dc}a=2\pi n\hbar/T$ for an integer $n$.
In scaled units, this equation becomes precisely the
``phase-locking'' condition
$\langle\Fsc\rangle=n\Omega$; intuitively, this
corresponds to resonant photon absorption between Stark
ladder levels separated by an integer multiple of the
photon energy.
Our results demonstrate that this picture generally continues
to be valid with the inclusion of scattering, when the Stark
levels broaden and the ``phase-locking'' condition is only
approximately satisfied.  In our situation of an SSL without external
bias, the self-consistent field leads to the spontaneous
formation of a Stark ladder with suitable spacing.

We conclude with two comments. First, we can estimate the size of
the spontaneously generated current corresponding to present
experimental parameters. For $\Omega=5.7$ THz
($\Omega/\omega_c\approx 0.2$) and
for the typical SSL parameters and dissipation constants
mentioned above, we find that
$I_{dc}\equiv(e N S \Delta a/ 2\hbar)\langle v\rangle\simeq 28\mu$A
$(\langle v\rangle\approx 0.002)$
at field strength $E_0\ge E_0^{(cr)}\approx 4$ kV/cm
($\omega_s^{(cr)}/\omega_c\approx 1$),
which should be within the parameter
range of current experiments
\cite{guimaraes,ignatov1,keay,unterrainer}.
Second, we should distinguish our results from a recent
study by Ignatov {\it et al.} \cite{ignatov2}, who
pointed out that absolute negative conductance in the
voltage-current characteristic of an SSL irradiated by an ac field of
frequency $\Omega$
could cause an instability of the zero bias-zero current state,
resulting in the switching to a new state
with dc voltage per SSL period close to $\hbar\Omega/e$.
These studies apply to a {\it biased} SSL operating in the high
frequency regime, in which there can be no chaos.
In contrast, our studies apply to an {\it unbiased} SSL
at lower frequencies and lower amplitudes of the ac field
and in a regime where chaotic behavior may occur.

We are grateful to Predrag Cvitanovi\'c, Alan Luther,
Sergei Turovets, and Boris Chirikov for discussions and
especially to Gennady Berman for discussions and on-going collaborations in
related work.
This work was partially supported by NATO(93-1602) and
benefited from computational support from NCSA.
K.N.A. acknowledges the support of INTAS(94-2058) and KRSF(6F0030),
E.H.C. that of USDoEd-P200A40532, and
J.C.M. that of USNSF-REU-PHYS93-22320.
\begin{figure}
\caption{Projection of the half-period stroboscopic
phase portrait on the velocity -- self-consistent field plane for
asymmetric chaos at $\Omega/\omega_c=0.47$, $\omega_s/\omega_c=0.78$,
$\gamma_v/\omega_c=0.1$, $\gamma_\varepsilon/\omega_c=\alpha/\omega_c=0.01$. }
\label{fig1}
\end{figure}
\begin{figure}
\caption{A plot of regions of symmetry-broken motion in the
external field frequency($\Omega$)-external field strength($\omega_s$) plane. }
\label{fig2}
\end{figure}
\begin{figure}
\caption{Dependence of the spontaneously generated bias
$\langle\Fsc\rangle/\omega_c$ on the external field frequency
$\Omega/\omega_c$ at $\omega_s/\omega_c=1.5$, $\gamma_v/\omega_c=0.1$,
$\gamma_\varepsilon/\omega_c=\alpha/\omega_c=0.01$ }
\label{fig3}
\end{figure}

\begin{references}
\bibitem{remark}
The effect of symmetry-breaking is very robust to changes of the dissipation 
rates: We observe it in the range $\gamma_v/\omega_c=0.01--1$.
The region of existence of chaotic attractors
in the $\Omega - \omega_s$ plane shrinks with increasing
dissipation rates.
\bibitem{note}
Compare in particular our Fig.~\ref{fig3}
and Fig 1 from \cite{d'humieres}.
\end{references}
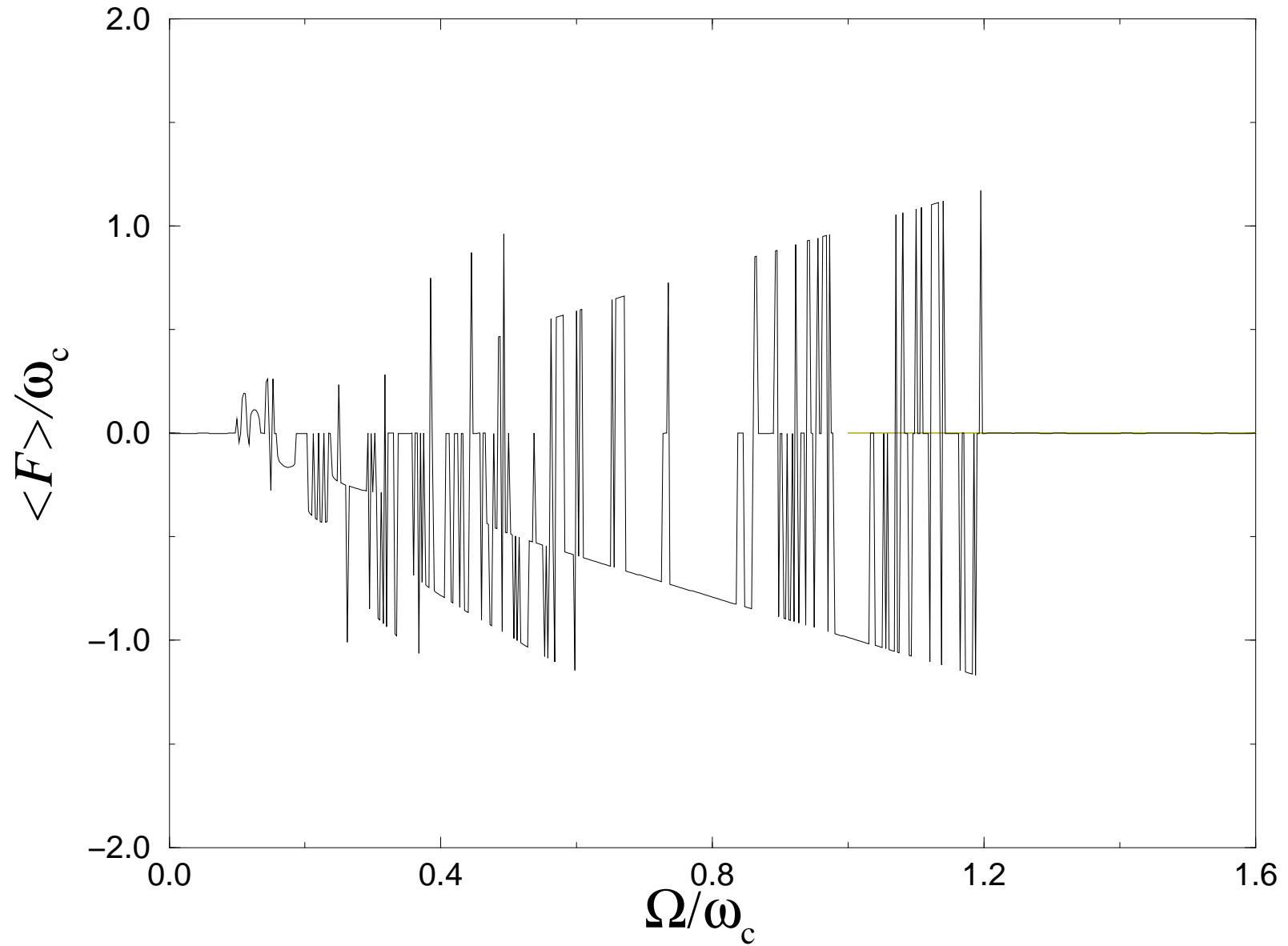
\end{document}